\documentclass[12pt]{article}
\usepackage{graphicx}

\newcommand{\beq}{\begin{equation}}
\newcommand{\eeq}{\end{equation}}
\newcommand{\beqa}{\begin{eqnarray}}
\newcommand{\eeqa}{\end{eqnarray}}
\newcommand{\non}{\nonumber}
\newcommand{\lab}{\label}
\newcommand{\bra}{\langle}
\newcommand{\ket}{\rangle}

\begin{document}

\title{Higher order perturbation theory for decoherence
in Grover's algorithm}

\author{Hiroo Azuma\thanks{On leave from Canon Inc., 5-1,
Morinosato-Wakamiya, Atsugi-shi, Kanagawa, 243-0193, Japan.}
\\
{\small Research Center for Quantum Information Science,}\\
{\small Tamagawa University Research Institute,}\\
{\small 6-1-1 Tamagawa-Gakuen, Machida-shi, Tokyo 194-8610, Japan}\\
{\small E-mail: h-azuma@lab.tamagawa.ac.jp}}

\date{July 29, 2005}

\maketitle

\begin{abstract}
In this paper, we study decoherence in Grover's
quantum search algorithm
using a perturbative method.
We assume that each two-state system (qubit)
that belongs to a register
suffers a phase flip error ($\sigma_{z}$ error)
with probability $p$ independently at every step in the algorithm,
where $0\leq p\leq 1$.
Considering an $n$-qubit density operator
to which Grover's iterative operation is applied $M$ times,
we expand it in powers of $2Mnp$ and derive its matrix element
order by order under the large-$n$ limit.
[In this large-$n$ limit,
we assume $p$ is small enough, so that $2Mnp$ can take
any real positive value or zero.
We regard $x\equiv 2Mnp(\geq 0)$ as a perturbative parameter.]
We obtain recurrence relations between terms
in the perturbative expansion.
By these relations, we compute higher orders
of the perturbation efficiently,
so that we extend the range of the perturbative parameter
that provides a reliable analysis.
Calculating the matrix element
numerically by this method,
we derive the maximum value of the perturbative parameter
$x$ at which the algorithm finds a correct item
with a given threshold of probability
$P_{\mbox{\scriptsize th}}$ or more.
(We refer to this maximum value of $x$ as $x_{\mbox{\scriptsize c}}$,
a critical point of $x$.)
We obtain a curve of $x_{\mbox{\scriptsize c}}$
as a function of $P_{\mbox{\scriptsize th}}$
by repeating this numerical calculation
for many points of
$P_{\mbox{\scriptsize th}}$
and find the following facts:
a tangent of the obtained curve at $P_{\mbox{\scriptsize th}}=1$
is given by
$x=(8/5)(1-P_{\mbox{\scriptsize th}})$,
and we have
$x_{\mbox{\scriptsize c}}>-(8/5)\log_{e}P_{\mbox{\scriptsize th}}$
near $P_{\mbox{\scriptsize th}}=0$.
\end{abstract}

\section{Introduction}
\lab{section-introduction}
Many researchers think that decoherence is one of the most serious
difficulties in realizing the quantum computation
\cite{Zurek,Unruh,Palma,Chuang}.
The decoherence
is caused by interaction between the quantum computer
and its environment.
The interaction
lets the state of the computer become correlated
with the state of the environment.
Consequently, some of information of the quantum computer leaks
into the environment.
This process causes errors in the state of the quantum computer,
and as a result,
the probability that the quantum algorithm gives
the right answer decreases.
To overcome this problem,
quantum error-correcting codes are proposed
\cite{Shor,Steane,Calderbank}.

Not only for practical purposes
but also for theoretical interests,
an important question is
how robust the quantum algorithm is against this disturbance.
If we know the upper bound of the error rate that allows the quantum
computer to obtain a solution with a certain probability or more,
this bound is useful for us to design quantum gates.

Grover's algorithm is considered to be
an efficient amplitude-amplification process
for quantum states.
Thus it is often called a search
algorithm \cite{Grover,Boyer}.
By applying the same unitary transformation to the state in iteration
and gradually amplifying an amplitude of one basis vector
that an oracle indicates,
Grover's algorithm picks it up from a uniform superposition of
$2^{n}$ basis vectors with a certain probability in $O(2^{n/2})$ steps.
In view of computational time
(the number of queries for the oracle),
the efficiency of Grover's algorithm is proved
to be optimal \cite{Ambainis}.

In Ref.~\cite{Azuma},
we study decoherence in Grover's algorithm
with a perturbative method.
We consider the following simple model.
First,
we assume that we search $|0...0\ket$
from the uniform superposition of $2^{n}$ logical basis vectors
$\{|x\ket:x\in\{0,1\}^{n}\}$ by Grover's algorithm.
This assumption simplifies the iterative transformation.
Second, we assume that each qubit of the register
interacts with the environment independently
and suffers a phase damping,
which causes a phase flip error
($\sigma_{z}$ error)
with probability $p$
and does nothing with probability $(1-p)$ to the qubit.
In this model,
we expand an $n$-qubit density operator
to which Grover's iterative operation is applied $M$ times
in powers of $2Mnp$.
Then, we take the large-$n$ limit,
so that we can simplify each order term
of the expansion of the density operator
and we obtain its asymptotic form.

In this large-$n$ limit, we assume $p$ is small enough,
so that $2Mnp$ can take any real positive value or zero.
We regard $x\equiv 2Mnp(\geq 0)$ as a perturbative parameter.
We can interpret $x=2Mnp$ as the expected number of phase
flip errors
($\sigma_{z}$ errors)
that occur during the running time of computation.
In Ref.~\cite{Azuma},
we give a formula for deriving an asymptotic form
of an arbitrary order term of the perturbative expansion.
However, this formula includes a complicated multiple integral
and the number of terms in its integrand increases exponentially.
Because of these difficulties,
we obtain explicit asymptotic forms
only up to the fifth-order term.

In this paper, using recurrence relations between terms
of the perturbative expansion,
we develop a method for computing higher order terms efficiently.
By this method, we derive an explicit form of the density matrix
of the disturbed quantum computer up to the $39$th-order term
with the help of a computer algebra system.
(In actual fact, we use {\it Mathematica} for this derivation.)
Because we consider the higher order perturbation,
we can greatly extend the range of the perturbative parameter
that provides a reliable analysis,
compared with our previous work in Ref.~\cite{Azuma}.
Calculating the matrix element up to the $39$th-order term
numerically
from the form obtained by the computer algebra system,
we derive the maximum value of the perturbative parameter
$x$ at which the algorithm finds a correct item
with a given threshold of probability $P_{\mbox{\scriptsize th}}$ or more.
(We refer to this maximum value of $x$ as $x_{\mbox{\scriptsize c}}$,
a critical point of $x$.)

Grover's algorithm can find the correct item
by less than $(\pi/4)\sqrt{2^{n}}$ steps
with given probability
$P_{\mbox{\scriptsize th}}$ or more under no decoherence ($p=0$).
When we fix
$P_{\mbox{\scriptsize th}}$,
the number of iterations that we need
increases as the decoherence becomes stronger ($p$ becomes larger).
Finally we never detect the correct item with $P_{\mbox{\scriptsize th}}$
or more for $p>p_{\mbox{\scriptsize c}}$.
Thus, we can think $p_{\mbox{\scriptsize c}}$ to be a critical point
for $P_{\mbox{\scriptsize th}}$.
($p_{\mbox{\scriptsize c}}$ depends on $P_{\mbox{\scriptsize th}}$.)
However, we actually obtain
$x_{\mbox{\scriptsize c}}=x_{\mbox{\scriptsize c}}(P_{\mbox{\scriptsize th}})$
for the perturbative parameter $x=2Mnp$
instead of
$p_{\mbox{\scriptsize c}}=p_{\mbox{\scriptsize c}}(P_{\mbox{\scriptsize th}})$.
From the relation
$x_{\mbox{\scriptsize c}}=x_{\mbox{\scriptsize c}}(P_{\mbox{\scriptsize th}})$,
we can draw a phase diagram
as shown in Fig.~\ref{pthxc_explain}.
The diagram consists of two domains.
One is where the quantum algorithm is effective
and the other is where it is not effective.

\begin{figure}
\begin{center}
\includegraphics[scale=1.0]{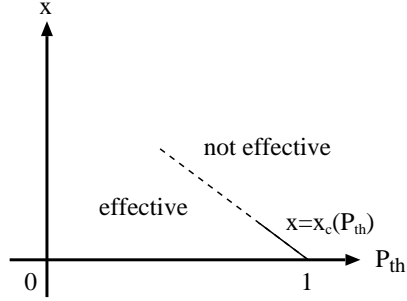}
\end{center}
\caption{A schematic representation of
$x_{\mbox{\scriptsize c}}$ as a function
of $P_{\mbox{\scriptsize th}}$.
$P_{\mbox{\scriptsize th}}$ is a threshold of probability.
$x$ represents both the perturbative parameter
and the expected number of errors
during the running time of computation.
$x_{\mbox{\scriptsize c}}$ is a critical point of $x$.
Both $P_{\mbox{\scriptsize th}}$ and $x$
are dimensionless.
We can easily obtain $x_{\mbox{\scriptsize c}}=0$
for $P_{\mbox{\scriptsize th}}=1$.
This fact is included in the above schematic graph.
The above graph represents a phase diagram that consists of two domains.
One domain is where the quantum algorithm is effective
and the other domain is where it is not effective.}
\lab{pthxc_explain}
\end{figure}

Figures~\ref{pthxc_linear} and \ref{pthxc_log}
represent a curve of
$x_{\mbox{\scriptsize c}}
=x_{\mbox{\scriptsize c}}(P_{\mbox{\scriptsize th}})$
obtained by repeating the numerical calculation
of $x_{\mbox{\scriptsize c}}$
for many points of $P_{\mbox{\scriptsize th}}$.
In Fig.~\ref{pthxc_linear}, we use a linear scale on both
horizontal and vertical axes.
We prove later that a tangent of the curve
$x=x_{\mbox{\scriptsize c}}(P_{\mbox{\scriptsize th}})$
at $P_{\mbox{\scriptsize th}}=1$ is given by
$x=(8/5)(1-P_{\mbox{\scriptsize th}})$.
In Fig.~\ref{pthxc_log}, we use log and  linear scales
on the horizontal and vertical axes, respectively. 
We observe
$x_{\mbox{\scriptsize c}}>-(8/5)\log_{e}P_{\mbox{\scriptsize th}}$
near $P_{\mbox{\scriptsize th}}=0$ from this figure.

\begin{figure}
\begin{center}
\includegraphics[scale=1.0]{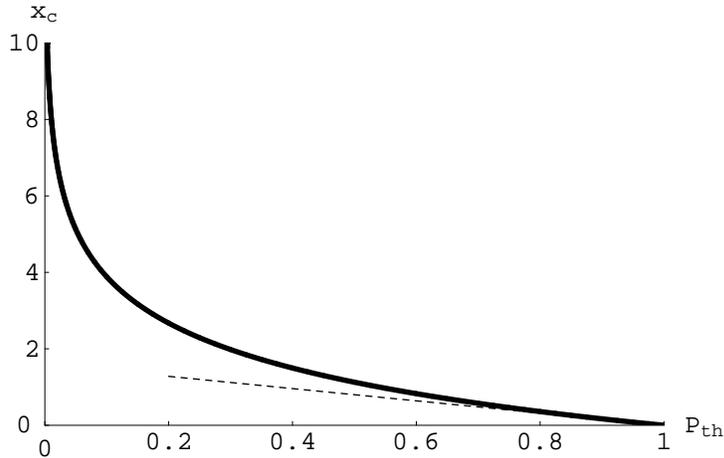}
\end{center}
\caption{$x_{\mbox{\scriptsize c}}$ as a function of
$P_{\mbox{\scriptsize th}}$.
[A thick solid curve represents
$x=x_{\mbox{\scriptsize c}}(P_{\mbox{\scriptsize th}})$.]
$P_{\mbox{\scriptsize th}}$ is a threshold of probability.
$x_{\mbox{\scriptsize c}}$ is a critical point of $x$
(the perturbative parameter).
We use a linear scale on both horizontal and vertical axes.
The data are obtained by repeating numerical calculation
of $x_{\mbox{\scriptsize c}}$
for many points of $P_{\mbox{\scriptsize th}}$.
Because $x=x_{\mbox{\scriptsize c}}(P_{\mbox{\scriptsize th}})$
shows a sharp divergence at $P_{\mbox{\scriptsize th}}=0$,
we start calculation of $x_{\mbox{\scriptsize c}}$
from $P_{\mbox{\scriptsize th}}=1$.
While we are going from $P_{\mbox{\scriptsize th}}=1$
toward $P_{\mbox{\scriptsize th}}=0$,
we make a finite difference of $P_{\mbox{\scriptsize th}}$
smaller gradually.
(We put $\Delta P_{\mbox{\scriptsize th}}=5.0\times 10^{-4}$
around $P_{\mbox{\scriptsize th}}=1$
and $\Delta P_{\mbox{\scriptsize th}}=5.0\times 10^{-7}$
around $P_{\mbox{\scriptsize th}}=3.7\times 10^{-3}$.)
A thin dashed line represents a tangent of
$x=x_{\mbox{\scriptsize c}}(P_{\mbox{\scriptsize th}})$
at $P_{\mbox{\scriptsize th}}=1$.}
\lab{pthxc_linear}
\end{figure}

\begin{figure}
\begin{center}
\includegraphics[scale=1.0]{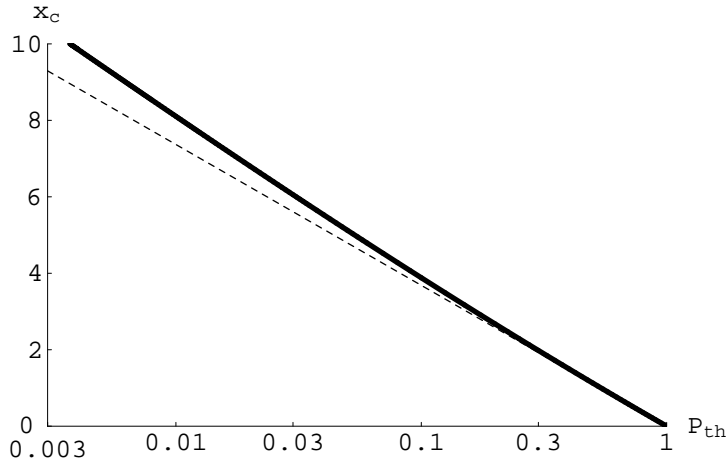}
\end{center}
\caption{$x_{\mbox{\scriptsize c}}$ as a function of
$P_{\mbox{\scriptsize th}}$.
[A thick solid curve represents
$x=x_{\mbox{\scriptsize c}}(P_{\mbox{\scriptsize th}})$.]
We use log and linear scales on the horizontal and vertical axes,
respectively.
In this figure, we use the same data of
$x=x_{\mbox{\scriptsize c}}(P_{\mbox{\scriptsize th}})$
as in Fig.~\ref{pthxc_linear}.
A thin dashed line represents
$x=-(8/5)\log_{e}P_{\mbox{\scriptsize th}}$.
We observe
$x_{\mbox{\scriptsize c}}>-(8/5)\log_{e}P_{\mbox{\scriptsize th}}$
near $P_{\mbox{\scriptsize th}}=0$.}
\lab{pthxc_log}
\end{figure}

Here, we mention that we can investigate our model
by Monte Carlo simulations, as well.
In fact, we compare results obtained by our perturbative method
with results obtained by Monte Carlo simulations
in Figs.~\ref{px-fixtheta} and \ref{ptheta-fixx}
in Sec.~\ref{section-numerical-calculations},
and we confirm that they are consistent.
From these analyses, we conclude that our perturbative method
is valid in a certain range of the perturbative parameter.

However, the Monte Carlo simulation method has
some difficulties for investigating our model.
First, the execution time of computation increases exponentially
in $n$ (the number of qubits).
We have to always come up against this problem
when we simulate a process of a quantum computer with a classical computer.
Second, the Monte Carlo simulation method is not suitable
for obtaining a variation of a physical quantity
as a function of some parameters,
because we carry out each simulation with fixing parameters
such as the error rate $p$ and the threshold
of probability $P_{\mbox{\scriptsize th}}$.
Thus we prefer our perturbative method to the Monte Carlo simulation method
for computing $x_{\mbox{\scriptsize c}}$ (the critical point of $x$)
that is obtained by evaluating the probability
of detecting a correct answer
as a function of $x$ and $P_{\mbox{\scriptsize th}}$.

A related result is obtained
in the study of the accuracy of quantum gates
by Bernstein and Vazirani \cite{Bernstein},
and Preskill \cite{Preskill}.
They consider a quantum circuit where each quantum gate has
a constant error because of inaccuracy.
Thus, it is an error of a unitary transformation and
it never causes dissipation of information
from the quantum computer to its environment.
They estimate inaccuracy $\epsilon$
for which the quantum algorithm is effective
under the fixed number of time steps $T$,
and obtain $2T\epsilon< 1-P_{\mbox{\scriptsize th}}$,
where $0\leq\epsilon\ll 1$.
If we regard $p/2$ as inaccuracy $\epsilon$
and $2Mn$ as the number of whole steps in the algorithm $T$,
it is similar to our observation that
$x_{\mbox{\scriptsize c}}
=2Mnp
\simeq
(8/5)(1-P_{\mbox{\scriptsize th}})$
near $P_{\mbox{\scriptsize th}}=1$,
except for a factor.

Barenco {\it et al}. study the approximate
quantum Fourier transformation (AQFT)
and its decoherence \cite{Barenco}.
Although their motivation is slightly different from
Refs.~\cite{Bernstein,Preskill},
we can think their model to be the quantum Fourier transformation (QFT)
with inaccurate gates.
They confirm that AQFT can give a performance
that is not much worse than the QFT.

This article is organized as follows.
In Sec.~\ref{section-model-and-perturbation},
we describe our model and perturbation theory
defined in our previous work \cite{Azuma}.
In Sec.~\ref{section-recurrence-relation},
we give recurrence relations
between terms of the perturbative expansion.
We develop a method for calculating higher order perturbation
efficiently with these relations.
In Sec.~\ref{section-numerical-calculations},
we carry out numerical calculations of the matrix element
of the density operator by the efficient method obtained
in Sec.~\ref{section-recurrence-relation}.
Moreover, we investigate the critical point
$x_{\mbox{\scriptsize c}}$
and obtain the phase diagram shown in Figs.~\ref{pthxc_linear}
and \ref{pthxc_log}.
In Sec.~\ref{section-discussions}, we give brief discussions.
In Appendix~\ref{appendix-proof-F-h-Theta},
we give a proof of an equation which appears
in Sec.~\ref{section-recurrence-relation}.

\section{Model and perturbation theory}
\label{section-model-and-perturbation}
In this section, we first describe a model that we analyze.
It is a quantum process of Grover's algorithm
under a phase damping at every iteration.
Second, we formulate a perturbation theory for this model.

\subsection{Model}
\lab{subsection-model}
First of all, we give a brief review of Grover's algorithm \cite{Grover}.
Starting from the $n$-qubit uniform superposition
of logical basis vectors,
\beq
W|0...0\ket
=
\frac{1}{\sqrt{2^{n}}}
\sum_{x\in\{0,1\}^{n}}|x\ket
\quad\quad
\mbox{for $n\geq 2$},
\lab{n-qubit-uniform-superposition}
\eeq
Grover's algorithm gradually amplifies the amplitude
of a certain basis vector $|x_{0}\ket$
that a quantum oracle indicates,
where $x_{0}\in\{0,1\}^{n}$.
The operator $W$ in Eq.~(\ref{n-qubit-uniform-superposition})
is an $n$-fold tensor product of a one-qubit unitary transformation
and given by $W=H^{\otimes n}$.
The operator $H$ is called Hadamard transformation
and represented by the following matrix,
\beq
H=\frac{1}{\sqrt{2}}
\left(
\begin{array}{cc}
1 & 1  \\
1 & -1
\end{array}
\right),
\eeq
where we use the orthonormal basis $\{|0\ket,|1\ket\}$
for this matrix representation.
The quantum oracle can be regarded as a black box
and actually it is a quantum gate
that shifts phases of logical basis vectors as
\beq
R_{x_{0}}:
\quad
\left\{
\begin{array}{llll}
|x_{0}\ket & \rightarrow & -|x_{0}\ket & \\
|x\ket     & \rightarrow & |x\ket & \mbox{for $x\neq x_{0}$}
\end{array}
\right. ,
\lab{selective-phase-shift-operator}
\eeq
where $x_{0},x \in\{0,1\}^{n}$.
[We note that all operators (quantum gates)
in Grover's algorithm are unitary.
Thus,
$H^{\dagger}=H^{-1}$, $W^{\dagger}=W^{-1}$,
$R_{x_{0}}^{\dagger}=R_{x_{0}}^{-1}$,
and so on.]

To let probability of observing $|x_{0}\ket$
be greater than a certain value ($1/2$, for example),
we repeat the following procedure $O(\sqrt{2^{n}})$ times:
\begin{enumerate}
\item Apply $R_{x_{0}}$ to the $n$-qubit state.
\item Apply $D=WR_{0}W$ to the $n$-qubit state.
\end{enumerate}
$R_{0}$ is a selective phase-shift operator,
which multiplies a factor $(-1)$ to $|0...0\ket$
and does nothing to the other basis vectors,
as defined in Eq.~(\ref{selective-phase-shift-operator}).
$D$ is called the inversion-about-average operation.

From now on,
we assume that
we amplify an amplitude of $|0...0\ket$.
From this assumption,
we can write an operation iterated in the algorithm as
\beq
DR_{0}=(WR_{0}W)R_{0}.
\eeq
After repeating this operation $M$ times
from the initial state $W|0\ket$($=W|0...0\ket$),
we obtain the state $(WR_{0})^{2M}W|0\ket$.
(We often write $|0\ket$
as an abbreviation of the $n$-qubit state
$|0...0\ket$ for a simple notation.)

Next, we think about the decoherence.
In this paper, we consider the following
one-qubit phase damping \cite{Huelga,Nielsen}:
\beq
\rho
\rightarrow
\rho'=p\sigma_{z}\rho\sigma_{z}+(1-p)\rho
\quad\quad
\mbox{for $0\leq p \leq 1$},
\lab{one-qubit-phase-error}
\eeq
where $\rho$ is an arbitrary one-qubit density operator.
$\sigma_{z}$ is one of the Pauli matrices
and given by
\beq
\sigma_{z}
=
\left(
\begin{array}{cc}
1 & 0  \\
0 & -1
\end{array}
\right) ,
\eeq
where we use the orthonormal basis $\{|0\ket,|1\ket\}$
for this matrix representation.
For simplicity,
we assume that
the phase damping of Eq.~(\ref{one-qubit-phase-error})
occurs in each qubit of the register independently
before every $R_{0}$ operation during the algorithm.
This implies that
each qubit interacts with its own environment independently.

Here,
we add some notes.
First, because $R_{0}\in U(2^{n})$ is
applied to all $n$ qubits
and
$H\in U(2)$ is applied to only one qubit,
we can imagine that
the realization of $R_{0}$
is more difficult than that of $W=H^{\otimes n}$.
Hence, we assume that the phase damping occurs only before $R_{0}$.
Second,
although we assume a very simple decoherence defined
in Eq.~(\ref{one-qubit-phase-error}),
we can think of other complicated disturbances.
For example, we can consider decoherence
caused by an interaction
between the environment and two qubits
and it may occur with a probability of $O(p^{2})$.
In this paper,
we do not assume such complicated disturbances.

\subsection{Perturbation theory}
\lab{subsection-perturbation-theory}
Let $\rho^{(M)}$ be the density operator obtained
by applying Grover's iteration $M$ times
to the $n$-qubit initial state $W|0\ket$.
The decoherence defined in Eq.~(\ref{one-qubit-phase-error})
occurs $2Mn$ times in $\rho^{(M)}$.
We can expand $\rho^{(M)}$ in powers
of $p$ and $(1-p)$ as follows:
\beqa
\rho^{(M)}
&=&
(1-p)^{2Mn}T_{0}^{(M)}
+(1-p)^{2Mn-1}p T_{1}^{(M)}
+... \non \\
&=&
\sum_{k=0}^{2Mn}(1-p)^{2Mn-k}p^{k} T_{k}^{(M)},
\lab{definition-(1-p)-p-expanded-density-operator}
\eeqa
where $\{T_{k}^{(M)}\}$ are given by
\beqa
T_{0}^{(M)}
&=&
(WR_{0})^{2M}W|0\ket\bra 0|W(R_{0}W)^{2M},
\lab{definition-T0M} \\
T_{1}^{(M)}
&=&
\sum_{i=1}^{n}
\sum_{l=0}^{2M-1}
(WR_{0})^{2M-l}\sigma_{z}^{(i)}(WR_{0})^{l}W|0\ket
\bra 0|W(R_{0}W)^{l}\sigma_{z}^{(i)}(R_{0}W)^{2M-l},
\lab{definition-T1M} \\
T_{2}^{(M)}
&=&
\sum_{i=1}^{n}
\sum_{{j=1}\atop{i<j}}^{n}
\sum_{l=0}^{2M-1}
(WR_{0})^{2M-l}\sigma_{z}^{(i)}\sigma_{z}^{(j)}
(WR_{0})^{l}W|0\ket\bra \mbox{H.c.}| \non \\
&&
+\sum_{i=1}^{n}
\sum_{j=1}^{n}
\sum_{l=0}^{2M-1}
\sum_{m=1}^{2M-l-1}
(WR_{0})^{2M-l-m}\sigma_{z}^{(i)}(WR_{0})^{m}\sigma_{z}^{(j)}
(WR_{0})^{l}W|0\ket \non \\
&&\quad\quad
\times\bra\mbox{H.c.}|,
\lab{definition-T2M}
\eeqa
and so on,
$\sigma_{z}^{(i)}$ represents the operator applied to the $i$th
qubit for $1\leq i\leq n$,
and $\bra \mbox{H.c.}|$ represents
a Hermitian conjugation of the ket vector
on its left side.
(Here, we note $W^{\dagger}=W$, $R_{0}^{\dagger}=R_{0}$,
and $\sigma_{z}^{(i)\dagger}=\sigma_{z}^{(i)}$.)
We can regard $T_{k}^{(M)}$ as a density operator
whose trace is not normalized.
It represents the sum of states
where $k$ errors occur
during the iteration of $M$ operations.

On the other hand,
from Eq.~(\ref{definition-(1-p)-p-expanded-density-operator}),
we can expand $\rho^{(M)}$
in powers of $p$ as follows:
\beqa
\rho^{(M)}
&=&
\rho_{0}^{(M)}
+2Mnp\rho_{1}^{(M)}
+\frac{1}{2}(2Mn)(2Mn-1)p^{2}\rho_{2}^{(M)}
+... \non \\
&=&
\sum_{k=0}^{2Mn}
{2Mn\choose{k}}
p^{k}\rho_{k}^{(M)},
\lab{density-operator-expansion-p-powers}
\eeqa
where
\beqa
\rho_{0}^{(M)}
&=&
T_{0}^{(M)}, \non \\
\rho_{1}^{(M)}
&=&
-T_{0}^{(M)}+\frac{T_{1}^{(M)}}{2Mn}, \non \\
\rho_{k}^{(M)}
&=&
(-1)^{k}\sum_{j=0}^{k}(-1)^{j}
{2Mn\choose{j}}^{-1}
{k\choose{j}}
T_{j}^{(M)}
\quad\quad
\mbox{for $k=0,1,...,2Mn$}.
\lab{Definitions-p-expanded-density-operators-0123}
\eeqa
Here, let us take the limit of an infinite number of qubits
(the large-$n$ limit).
We assume that we can take very small $p$,
so that $2Mnp$ can be an arbitrary real positive value or zero.
If $2Mnp$ is small enough, we can consider
$x\equiv 2Mnp(\geq 0)$ to be a perturbative parameter and the series of
Eq.~(\ref{density-operator-expansion-p-powers})
to be a perturbative expansion.

Under this limit,
we derive an asymptotic form of $\bra 0|\rho^{(M)}|0\ket$. 
In the actual derivation, we take the limit of $n\rightarrow \infty$
with holding $x=2Mnp$ finite.
$\bra 0|\rho^{(M)}|0\ket$ is a probability that the quantum
computer finds a correct item after $M$ operations.
Because we divide $T_{j}^{(M)}$ by $(Mn)^{j}$
as in Eq.~(\ref{Definitions-p-expanded-density-operators-0123}),
an expectation value of $\rho_{k}^{(M)}$ 
can converge to a finite value in the limit $n\rightarrow \infty$
for $k=0,1,...,2Mn$.

With these preparations,
we will investigate the following physical quantities.
Let $P_{\mbox{\scriptsize th}}$ be a threshold of probability
for $0< P_{\mbox{\scriptsize th}}\leq 1$,
so that if the quantum computer finds a correct item
(in our model, it is $|0\ket$)
with probability $P_{\mbox{\scriptsize th}}$ or more,
we regard it effective,
and otherwise we do not consider it effective.
Then, we consider the least number of the operations
that we need to repeat for amplifying the probability of
observing $|0\ket$ to $P_{\mbox{\scriptsize th}}$ or more
for a given $p$.
We refer to it as
$M_{\mbox{\scriptsize th}}(p,P_{\mbox{\scriptsize th}})$.
[$M_{\mbox{\scriptsize th}}(p,P_{\mbox{\scriptsize th}})$
is the least number of $M$
that satisfies
$\bra 0|\rho^{(M)}|0\ket=P_{\mbox{\scriptsize th}}$
for a given $p$.]
As $p$ becomes larger with fixing $P_{\mbox{\scriptsize th}}$,
we can expect that
$M_{\mbox{\scriptsize th}}(p,P_{\mbox{\scriptsize th}})$
increases monotonically.
In the end,
we never observe $|0\ket$ at least
with a probability $P_{\mbox{\scriptsize th}}$
for a certain $p_{\mbox{\scriptsize c}}$ or more.
(Hence, $p_{\mbox{\scriptsize c}}$ depends on
$P_{\mbox{\scriptsize th}}$.)
Regarding $P_{\mbox{\scriptsize th}}$ as a threshold
for whether the quantum computer is effective or not,
we can consider $p_{\mbox{\scriptsize c}}$ to be a critical point.

In our perturbation theory,
we calculate physical quantities
using the dimensionless perturbative parameter $x=2Mnp$.
Thus, we take $M$ and $x$ for independent variables.
(In our original model defined in Sec.~\ref{subsection-model},
we take $M$ and $p$ for independent variables.)
We can define as well
$\tilde{M}_{\mbox{\scriptsize th}}(x,P_{\mbox{\scriptsize th}})$
that represents the least number of the operations iterated
for amplifying the probability of $|0\ket$ to $P_{\mbox{\scriptsize th}}$
for given $x$.
Furthermore, we also obtain $x_{\mbox{\scriptsize c}}$ or more
for which we can never detect $|0\ket$ at least
with probability $P_{\mbox{\scriptsize th}}$.

Next, we evaluate $\bra 0|\rho^{(M)}|0\ket$.
First, from simple calculation,
we obtain the unperturbed matrix element,
\beqa
\bra 0|\rho^{(M)}|0\ket_{p=0}
&=&\bra 0|T_{0}^{(M)}|0\ket \non \\
&=&\sin^{2}[(2M+1)\theta],
\lab{matrix-element-T0}
\eeqa
where
\beq
\sin\theta=\frac{1}{\sqrt{2^{n}}},
\quad\quad
\cos\theta=\sqrt{\frac{2^{n}-1}{2^{n}}}.
\lab{Boyer-parameter-theta}
\eeq
(This parameter $\theta$ is introduced by Boyer {\it et al}.
\cite{Boyer}.)
From Eq.~(\ref{matrix-element-T0}),
we notice the following facts.
If there is no decoherence ($p=0$),
we can amplify the probability of
observing $|0\ket$ to unity.
Taking large (but finite) $n$,
we obtain $\sin\theta\simeq\theta$
and $\theta\simeq 1/\sqrt{2^{n}}$,
and
we can observe $|0\ket$ with unit probability
after repeating Grover's operation
$M_{\mbox{\scriptsize max}}\simeq (\pi/4)\sqrt{2^{n}}$ times.

To describe the asymptotic forms of matrix elements,
we introduce the following notation.
Because $\bra 0|T_{0}^{(M)}|0\ket$ is a periodic function of $M$
and its period is about $\pi\sqrt{2^{n}}$
under the large-$n$ limit,
it is convenient for us to define a new variable
$\Theta =\lim_{n\rightarrow\infty}M\theta$ (radian).
Here, we give a formula for the asymptotic form of the $k$th
order of the matrix element under $n\rightarrow\infty$
for $k=1,2,...$.
(The derivation of this formula is given in
Secs.~4--7 and Appendix A of Ref.~\cite{Azuma}.)
Preparing an $k$-digit binary string
$\alpha=(\alpha_{1},...,\alpha_{k})\in\{0,1\}^{k}$,
we define the following $2^{k}$ terms:
\beqa
\lefteqn{|\tilde{{\cal T}}_{\alpha_{1},...,\alpha_{k}}
(\phi_{1},...,\phi_{k})|^{2}} \non \\
&=&
[
{\sin\brace{\cos}}_{\alpha_{1}}(2\phi_{1})
{\cos\brace{\sin}}_{\alpha_{2}}(2\phi_{2})
...
{\cos\brace{\sin}}_{\alpha_{k}}(2\phi_{k})
{\cos\brace{\sin}}_{\oplus_{s=1}^{k}\alpha_{s}}
(2[\Theta-\sum_{s=1}^{k}\phi_{s}])
]^{2} \non \\
&&
\quad\quad
\mbox{for $k=1,2,...$},
\lab{Diagrammatic-rule}
\eeqa
where
\beq
{f\brace{g}}_{\alpha}(x)
=
\left\{
\begin{array}{lll}
f(x) & \mbox{for $\alpha=0$} \\
g(x) & \mbox{for $\alpha=1$}
\end{array}
\right.,
\eeq
and $\oplus$ denotes the addition modulo $2$.
We notice that
the function of $\phi_{1}$ and the other functions of
$\phi_{2}$,..., $\phi_{k}$,
$\Theta-\sum_{s=1}^{k}\phi_{s}$
are different
(sine and cosine functions are put in reverse).
These terms are integrated as
\beqa
\lefteqn{\lim_{n\rightarrow \infty}
\frac{\bra 0|T_{k}^{(M)}|0\ket}{(Mn)^{k}}} \non \\
&=&
\frac{1}{\Theta^{k}}
\int_{0}^{\Theta}d\phi_{1}
\int_{0}^{\Theta-\phi_{1}}d\phi_{2}
...
\int_{0}^{\Theta-\phi_{1}-...-\phi_{k-1}}d\phi_{k} \non \\
&&\quad\quad
\times
\sum_{(\alpha_{1},...,\alpha_{k})\in\{0,1\}^{k}}
|\tilde{{\cal T}}_{\alpha_{1},...,\alpha_{k}}
(\phi_{1},...,\phi_{k})|^{2}.
\lab{Diagrammatic-rule-integral}
\eeqa

We can obtain the matrix elements as follows.
From Eq.~(\ref{matrix-element-T0}), we obtain
\beq
\lim_{n\rightarrow \infty}\bra 0|T_{0}^{(M)}|0\ket
=\sin^{2}2\Theta.
\lab{asymptotic-form-T0M}
\eeq
From Eqs.~(\ref{Diagrammatic-rule}) and (\ref{Diagrammatic-rule-integral}),
we obtain
\beqa
\lim_{n\rightarrow \infty}
\frac{\bra 0|T_{1}^{(M)}|0\ket}{Mn}
&=&
\frac{1}{\Theta}
\int_{0}^{\Theta}d\phi
\{
[\sin 2\phi\cos 2(\Theta-\phi)]^{2}
+
[\cos 2\phi\sin 2(\Theta-\phi)]^{2}
\} \non \\
&=&
\frac{1}{2}
-\frac{1}{4}\cos 4\Theta
-\frac{1}{16\Theta}\sin 4\Theta,
\lab{asymptotic-T1M-integral-form}
\eeqa
\beqa
\lim_{n\rightarrow \infty}
\frac{\bra 0|T_{2}^{(M)}|0\ket}{(Mn)^{2}}
&=&
\frac{1}{\Theta^{2}}
\int_{0}^{\Theta}d\phi
\int_{0}^{\Theta-\phi}d\varphi \non \\
&&\quad\quad
\times
\{[\sin 2\phi \cos 2\varphi \cos 2(\Theta-\phi-\varphi)]^{2} \non \\
&&\quad\quad
+[\cos 2\phi \cos 2\varphi \sin 2(\Theta-\phi-\varphi)]^{2} \non \\
&&
\quad\quad
+[\sin 2\phi \sin 2\varphi \sin 2(\Theta-\phi-\varphi)]^{2} \non \\
&&
\quad\quad
+[\cos 2\phi \sin 2\varphi \cos 2(\Theta-\phi-\varphi)]^{2}\} \non \\
&=&
\frac{1}{4}
-\frac{1}{16}\cos 4\Theta
-\frac{3}{64\Theta}\sin 4\Theta,
\lab{asymptotic-T2M-integral-form}
\eeqa
and so on.

The asymptotic form of the perturbative expansion of
the whole density matrix is given by
\beqa
\bra P\ket(\Theta,x)
&=&
\lim_{n\rightarrow\infty}
\bra 0|\rho^{(M)}|0\ket \non \\
&=&
C_{0}(\Theta)+C_{1}(\Theta)x+\frac{1}{2}C_{2}(\Theta)x^{2}+... \non \\
&=&
\sum_{k=0}^{\infty}C_{k}(\Theta)\frac{1}{k!}x^{k},
\lab{matrix-element-asymptotic-expansion}
\eeqa
where
\beqa
C_{0}(\Theta)&=&F_{0}(\Theta), \non \\
C_{1}(\Theta)&=&-F_{0}(\Theta)+\frac{1}{2}F_{1}(\Theta), \non \\
C_{k}(\Theta)
&=&
(-1)^{k}\sum_{j=0}^{k}
(-\frac{1}{2})^{j}
\frac{k!}{(k-j)!}
F_{j}(\Theta)
\quad\quad
\mbox{for $k=0,1,...$},
\lab{coefficients-matrix-element-asymptotic-expansion}
\eeqa
and
\beq
F_{k}(\Theta)
=\lim_{n\rightarrow \infty}
\frac{\bra 0|T_{k}^{(M)}|0\ket}{(Mn)^{k}}
\quad\quad
\mbox{for $k=0,1,...$}.
\lab{Definition-F-h-Theta}
\eeq
In Eq.~(\ref{matrix-element-asymptotic-expansion}),
the $k$th-order term is divided by $k!$,
so that we can expect the series $\bra P\ket(\Theta,x)$
to converge to a finite value for large $x$.

\section{Recurrence relations between order terms}
\label{section-recurrence-relation}
In this section, we obtain recurrence relations between order
terms of the perturbative series.
Using this result,
we develop a method for computing higher order terms efficiently.

When we compute
$\lim_{n\rightarrow \infty}\bra 0|T_{k}^{(M)}|0\ket/(Mn)^{k}$
for large $k$
from Eqs.~(\ref{Diagrammatic-rule}) and (\ref{Diagrammatic-rule-integral}),
we notice the following difficulties:
\begin{enumerate}
\item Equation~(\ref{Diagrammatic-rule-integral})
includes an $k$th-order integral.
\item Equation~(\ref{Diagrammatic-rule-integral})
includes $2^{k}$ terms being integrated.
\end{enumerate}
Even if we use a computer algebra system,
these troubles are serious.
(In Ref.~\cite{Azuma}, we obtain
$\lim_{n\rightarrow \infty}\bra 0|T_{k}^{(M)}|0\ket/(Mn)^{k}$
only up to $k=5$.)

To develop an efficient derivation of higher order terms,
we pay attention to the following relations:
\beq
F_{k}(\Theta)
=\lim_{n\rightarrow \infty}
\frac{\bra 0|T_{k}^{(M)}|0\ket}{(Mn)^{k}}
=\frac{f_{k}(\Theta)}{\Theta^{k}}
\quad\quad
\mbox{for $k=0,1,...$},
\lab{another-Definition-F-h-Theta}
\eeq
where
\beqa
f_{0}(\Theta)&=&\sin^{2}2\Theta, \lab{f-0-Theta-definition} \\
g_{0}(\Theta)&=&\cos^{2}2\Theta, \lab{g-0-Theta-definition} \\
f_{k}(\Theta)&=&
\int^{\Theta}_{0}d\phi
[f_{k-1}(\Theta-\phi)\cos^{2}2\phi
+g_{k-1}(\Theta-\phi)\sin^{2}2\phi], \lab{f-k-Theta-definition} \\
g_{k}(\Theta)&=&
\int^{\Theta}_{0}d\phi
[g_{k-1}(\Theta-\phi)\cos^{2}2\phi
+f_{k-1}(\Theta-\phi)\sin^{2}2\phi] \lab{g-k-Theta-definition} \\
&&\quad\quad
\mbox{for $k=1,2,...$.} \non
\eeqa
We can prove the above relations from
Eqs.~(\ref{Diagrammatic-rule}) and (\ref{Diagrammatic-rule-integral}).
Both Eqs.~(\ref{f-k-Theta-definition}) and (\ref{g-k-Theta-definition})
contain only first-order integrals.
Moreover, each of them contains only two terms being integrated.
Thus, we can compute $F_{0}(\Theta)$, $F_{1}(\Theta)$, ...
in that order efficiently from
Eqs.~(\ref{another-Definition-F-h-Theta}),
(\ref{f-0-Theta-definition}), (\ref{g-0-Theta-definition}),
(\ref{f-k-Theta-definition}), and (\ref{g-k-Theta-definition})
using a computer algebra system.
(In actual fact, we use {\it Mathematica}
for this derivation.)
Eqs.~(\ref{f-k-Theta-definition}) and (\ref{g-k-Theta-definition})
constitute a pair of recurrence formulas.

Here, we note some properties of $F_{k}(\Theta)$.
First, $f_{k}(\Theta)$ and $g_{k}(\Theta)$ are
analytic at any $\Theta$ for $k=0,1,...$.
In other words,
$f_{k}(\Theta)$ and $g_{k}(\Theta)$ have Taylor expansions
about any $\Theta_{0}$ which converge to
$f_{k}(\Theta)$ and $g_{k}(\Theta)$
in some neighborhood of $\Theta_{0}$ for $k=0,1,...$,
respectively.
We can prove these facts by mathematical induction
as follows.
To begin with,
both $f_{0}(\Theta)$ and $g_{0}(\Theta)$
are analytic at any $\Theta$
from Eqs.~(\ref{f-0-Theta-definition}) and
(\ref{g-0-Theta-definition}).
Next, we assume that $f_{k}(\Theta)$ and $g_{k}(\Theta)$ are
analytic at any $\Theta$ for some $k\in\{0,1,...\}$.
Then, $f_{k+1}(\Theta)$ and $g_{k+1}(\Theta)$ are
analytic at any $\Theta$
because they are integrals of functions made of the
sine and cosine functions,
$f_{k}(\Theta)$, and $g_{k}(\Theta)$,
as shown in Eqs.~(\ref{f-k-Theta-definition})
and (\ref{g-k-Theta-definition}).
Thus, by mathematical induction,
we conclude that
$f_{k}(\Theta)$ and $g_{k}(\Theta)$ are
analytic functions for $k=0,1,...$.

From Eq.~(\ref{another-Definition-F-h-Theta}),
we can obtain $F_{k}(\Theta)$
by dividing $f_{k}(\Theta)$ by $\Theta^{k}$
for $k=0,1,...$.
Thus, it is possible that $F_{k}(\Theta)$
diverges by marching off to infinity near $\Theta=0$.
However, in fact we can show
\beq
F_{k}(\Theta)
=\frac{f_{k}(\Theta)}{\Theta^{k}}
=\mbox{Const.}\Theta^{2}+O(\Theta^{4})
\quad\quad
\mbox{for $k=0,1,...$},
\lab{F-h-Theta-expansion}
\eeq
where Const. denotes some constant.
(We prove Eq.~(\ref{F-h-Theta-expansion})
in Appendix~\ref{appendix-proof-F-h-Theta}.)

\section{Numerical calculations}
\lab{section-numerical-calculations}
In this section, we carry out numerical calculations of
$\bra P\ket(\Theta,x)$ defined
in Eq.~(\ref{matrix-element-asymptotic-expansion})
using recurrence relations Eqs.~(\ref{f-k-Theta-definition})
and (\ref{g-k-Theta-definition}).
Moreover, we investigate the critical point $x_{\mbox{\scriptsize c}}$,
over which the quantum algorithm becomes ineffective
for the threshold probability $P_{\mbox{\scriptsize th}}$.

First of all, we need to derive an algebraic representation
of $\bra P\ket(\Theta,x)$.
We compute an explicit form of $\bra P\ket(\Theta,x)$ as follows.
First, using recurrence relations Eqs.~(\ref{f-k-Theta-definition})
and (\ref{g-k-Theta-definition}),
we derive $f_{k}(\Theta)$ and $g_{k}(\Theta)$.
Second, using Eq.~(\ref{another-Definition-F-h-Theta}),
we derive $F_{k}(\Theta)$ from $f_{k}(\Theta)$.
Next, using Eq.~(\ref{coefficients-matrix-element-asymptotic-expansion}),
we derive $C_{k}(\Theta)$ from $F_{k}(\Theta)$.
Finally, using Eq.~(\ref{matrix-element-asymptotic-expansion}),
we derive $\bra P\ket(\Theta,x)$ from $C_{k}(\Theta)$,
which is the $k$th-order term of the perturbative expansion.

In Ref.~\cite{Azuma}, we obtain an explicit form of the matrix element
only up to the fifth-order perturbation
[that is, $F_{5}(\Theta)$]
because we compute $F_{k}(\Theta)$
from Eq.~(\ref{Diagrammatic-rule-integral}) directly.
However, in this paper, we succeed in deriving an explicit form
of the matrix element up to the $39$th-order perturbation
[that is, $F_{39}(\Theta)$]
with the help of the computer algebra system
thanks to the recurrence relations Eqs.~(\ref{f-k-Theta-definition})
and (\ref{g-k-Theta-definition}).
[We do not write down the explicit forms of
$F_{3}(\Theta)$, ..., $F_{39}(\Theta)$ here
except for $F_{5}(\Theta)$,
because they are very complicated.]
By the method explained above, we derive the algebraic form of
$\bra P\ket(\Theta,x)$ up to the $39$th-order perturbation.

However, this explicit form of $\bra P\ket(\Theta,x)$
is not suitable for numerical calculation.
The reason is as follows.
Let us consider $F_{5}(\Theta)$ for example.
The explicit form of $F_{5}(\Theta)$ is given by
\beqa
F_{5}(\Theta)
&=&
\frac{1}{240}
+\frac{45+720\Theta^{2}-256\Theta^{4}}{1\;966\;080\Theta^{4}}\cos 4\Theta
\non \\
&&\quad\quad
-\frac{3+32\Theta^{2}+256\Theta^{4}}{524\;288\Theta^{5}}\sin 4\Theta.
\lab{F-5-Theta-explicit-form}
\eeqa
It is very difficult to evaluate the value of $F_{5}(\Theta)$
near $\Theta=0$ from Eq.~(\ref{F-5-Theta-explicit-form}) directly.
If we take the limit $\Theta\rightarrow 0$
in the second term of Eq.~(\ref{F-5-Theta-explicit-form}),
we obtain
\beq
\lim_{\Theta\rightarrow 0}
\frac{45+720\Theta^{2}-256\Theta^{4}}{1\;966\;080\Theta^{4}}\cos 4\Theta
=+\infty.
\eeq
However, taking the limit $\Theta\rightarrow 0$
in the third term of Eq.~(\ref{F-5-Theta-explicit-form}),
we obtain
\beqa
&&
\lim_{\Theta\rightarrow 0}
(-\frac{3+32\Theta^{2}+256\Theta^{4}}{524\;288\Theta^{5}})\sin 4\Theta
\non \\
&=&
-\lim_{\Theta\rightarrow 0}
(\frac{3+32\Theta^{2}+256\Theta^{4}}{131\;072\Theta^{4}})
\frac{\sin 4\Theta}{4\Theta} \non \\
&=&-\infty.
\eeqa
As explained above, to evaluate the value of $F_{5}(\Theta)$
near $\Theta=0$ from Eq.~(\ref{F-5-Theta-explicit-form}) directly,
we have to subtract one huge value from another huge value.
Thus, if we carry out this operation by computer,
an underflow error occurs and we cannot predict a result of the numerical
calculation at all.
[We show that $F_{k}(\Theta)$ is analytic at $\Theta=0$
and $\lim_{\Theta\rightarrow 0}F_{k}(\Theta)=0$
for $k=0,1,...$ in Eq.~(\ref{F-h-Theta-expansion}).
However, it is difficult to calculate $F_{5}(\Theta)$
numerically from Eq.~(\ref{F-5-Theta-explicit-form}).]

In fact, when $\Theta=1.0\times 10^{-7}$,
the second term of Eq.~(\ref{F-5-Theta-explicit-form})
is equal to $2.28881\times 10^{23}$
and the third term of Eq.~(\ref{F-5-Theta-explicit-form})
is equal to $-2.28881\times 10^{23}$
with assuming that the computer supports only six significant figures.
Hence, the sum of the second term and the third term
in Eq.~(\ref{F-5-Theta-explicit-form}) is equal to zero,
and only the first term of Eq.~(\ref{F-5-Theta-explicit-form}),
$1/240$, contributes to $F_{5}(\Theta)$ for $\Theta=1.0\times 10^{-7}$.
However, this numerical calculation is meaningless.
We can find such an underflow error in almost all the higher order terms,
$F_{3}(\Theta)$, $F_{4}(\Theta)$, $F_{5}(\Theta)$, ....

To avoid this trouble, we carry out the following procedure.
We expand the explicit form of $C_{k}(\Theta)$ in powers of $\Theta$
up to the $40$th-order term
and define $\bar{C}_{k}(\Theta)$ as the finite power series obtained
in the variable $\Theta$
for $k=0,1,...,39$.
[$C_{k}(\Theta)$ is originally an analytic function
and it has a Taylor expansion about $\Theta=0$.]
We substitute these $\{\bar{C}_{k}(\Theta):k=0,1,...,39\}$
for Eq.~(\ref{matrix-element-asymptotic-expansion}) and obtain
\beq
\bra \bar{P}\ket(\Theta,x)
=\sum_{k=0}^{39}\bar{C}_{k}(\Theta)\frac{1}{k!}x^{k}.
\lab{matrix-element-asymptotic-expansion-upto-39}
\eeq
We use this $\bra \bar{P}\ket(\Theta,x)$ for numerical calculation.
[$\bra \bar{P}\ket(\Theta,x)$ is a polynomial,
whose highest power in $\Theta$ is equal to $40$
and whose highest power in $x$ is equal to $39$.]

Here, we make some comments on our approximation method
for $C_{k}(\Theta)$.
In this paper, we use a polynomial of high degree for approximating
$C_{k}(\Theta)$.
The reasons for this choice are as follows:
(1) to obtain the Taylor series of $C_{k}(\Theta)$ is easy, and
(2) because we can calculate integrals and derivatives of polynomials
with ease,
$\bar{C}_{k}(\Theta)$ is suitable for applying Newton's method.
(We use Newton's method for calculating $x_{\mbox{\scriptsize c}}$
later.)
However, approximation with a polynomial of high degree sometimes
causes oscillations, and consequently errors of numerical calculation.
Pad{\'e} approximant method is effective in the treatment of this problem.
However, we do not use this method here,
because we have to carry out tough calculations for deriving
the Pad{\'e} approximants of $C_{k}(\Theta)$.

\begin{figure}
\begin{center}
\includegraphics[scale=1.0]{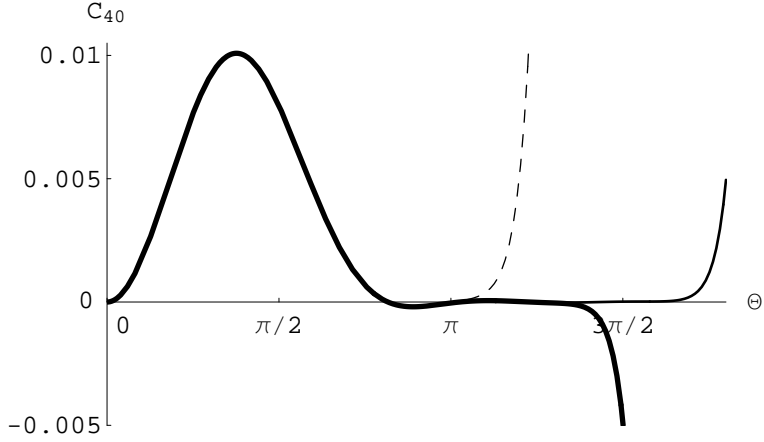}
\end{center}
\caption{$30$th, $40$th, and $50$th-order polynomials,
which we obtain as parts of Taylor series of $C_{40}(\Theta)$,
as functions of $\Theta$ for $0\leq\Theta\leq (9/5)\pi$.
A dashed line, a thick solid line, and a thin solid line represent
the $30$th, $40$th, and $50$th-order polynomials, respectively.}
\lab{C40-polynomial}
\end{figure}

We use a $40$th-order polynomial for approximating $C_{k}(\Theta)$
in this paper.
Figure~\ref{C40-polynomial} shows
$30$th, $40$th, and $50$th-order polynomials,
which we obtain as parts of Taylor series of $C_{40}(\Theta)$,
as functions of $\Theta$ for $0\leq\Theta\leq (9/5)\pi$.
A dashed line, a thick solid line, and a thin solid line represent
the $30$th, $40$th, and $50$th-order polynomials, respectively.
From Fig.~\ref{C40-polynomial}, we find that the $30$th, $40$th,
and $50$th-order polynomials start to diverge near
$\Theta=3.4$, $4.3$, and $5.3$(radian), respectively.
From this observation, we think the approximation of $C_{40}(\Theta)$
with the $40$th-order polynomial,
that is $\bar{C}_{40}(\Theta)$, to be valid
in the range of $0\leq\Theta\leq\pi$.
[Strictly speaking, this is not a rigorous proof but evidence
that the polynomial expansion up to the $40$th-order term is
sufficient for approximating $C_{k}(\Theta)$ for $k=0,1,...,39$
in the range of $0\leq\Theta\leq\pi$.]

To investigate the range of $x$ where our perturbative
approach is valid,
we need to estimate the $40$th-order perturbation.
From numerical calculation,
we obtain
\beq
0\leq |\frac{1}{40!}\bar{C}_{40}(\Theta)|\leq 1.24\times 10^{-50}
\lab{40th-order-perturbation-1}
\eeq
for $0\leq\Theta\leq\pi$.
(From now on, we limit $\Theta$ to $0\leq\Theta\leq\pi$
for our analysis,
because the approximation of $C_{k}(\Theta)$ with the $40$th-order
polynomial is reliable in this range,
as shown in Fig.~\ref{C40-polynomial}.)
Hence, if we limit $x$ to
$0\leq x \leq 10.0$,
the $40$th-order perturbation is bounded to
\beq
0\leq |\frac{1}{40!}\bar{C}_{40}(\Theta)x^{40}|\leq 1.24\times 10^{-10}.
\lab{40th-order-perturbation-2}
\eeq
[From now on, we write $\bra P\ket(\Theta,x)$ as
the approximate form $\bra \bar{P}\ket(\Theta,x)$
for convenience as far as
this naming does not create any confusion.]

Let us investigate $\bra P\ket(\Theta,x)$ obtained
in Eq.~(\ref{matrix-element-asymptotic-expansion-upto-39})
by numerical
calculations.
To confirm reliability of our perturbation theory,
we compare the obtained $\bra P\ket(\Theta,x)$ with results of Monte Carlo
simulations of our model in Figs.~\ref{px-fixtheta} and \ref{ptheta-fixx}.
In these simulations,
setting $n=9$ (nine qubits),
we fix $p$ and cause phase flip errors ($\sigma_{z}$ errors) at random
in each trial.
We take an average of
$\bra 0| \rho^{(M)}|0\ket_{p}$,
the probability of observing $|0\ket$ at the $M$th step
[$M=0,1,...,M_{\mbox{\scriptsize max}}(=17)$],
with $50{\,}000$ trials for each certain value of $p$.
[Because $(\pi/4)\sqrt{2^{9}}=17.7...$, we put
$M_{\mbox{\scriptsize max}}=17$.]

\begin{figure}
\begin{center}
\includegraphics[scale=1.0]{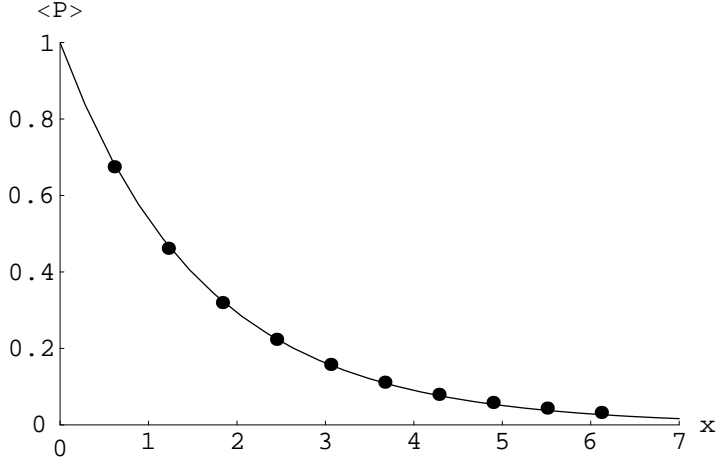}
\end{center}
\caption{$\bra P\ket(\Theta,x)$ as a function
of $x$ with fixing $\Theta=\Theta_{\mbox{\scriptsize max}}$,
where
$\Theta_{\mbox{\scriptsize max}}=
(1/2)(2M_{\mbox{\scriptsize max}}+1)\theta$,
$M_{\mbox{\scriptsize max}}=17$,
$\theta=\arcsin\sqrt{1/2^{n}}$, and $n=9$.
Both $\bra P\ket$ and $x$ are dimensionless.
A thin solid curve represents $\bra P\ket(\Theta,x)$
obtained by numerical calculation
up to the $39$th-order perturbation.
Black circles represent results obtained by Monte Carlo
simulations of the $n=9$ case (nine qubits)
with $M_{\mbox{\scriptsize max}}=17$.
Each circle is obtained for $x=2M_{\mbox{\scriptsize max}}np=306p$,
where $p$ is varied from $p=2.0\times 10^{-3}$ to
$p=2.0\times 10^{-2}$ at intervals of $\Delta p=2.0\times 10^{-3}$.
In these simulations, we make $50{\,}000$ trials for taking an average.}
\lab{px-fixtheta}
\end{figure}

\begin{figure}
\begin{center}
\includegraphics[scale=1.0]{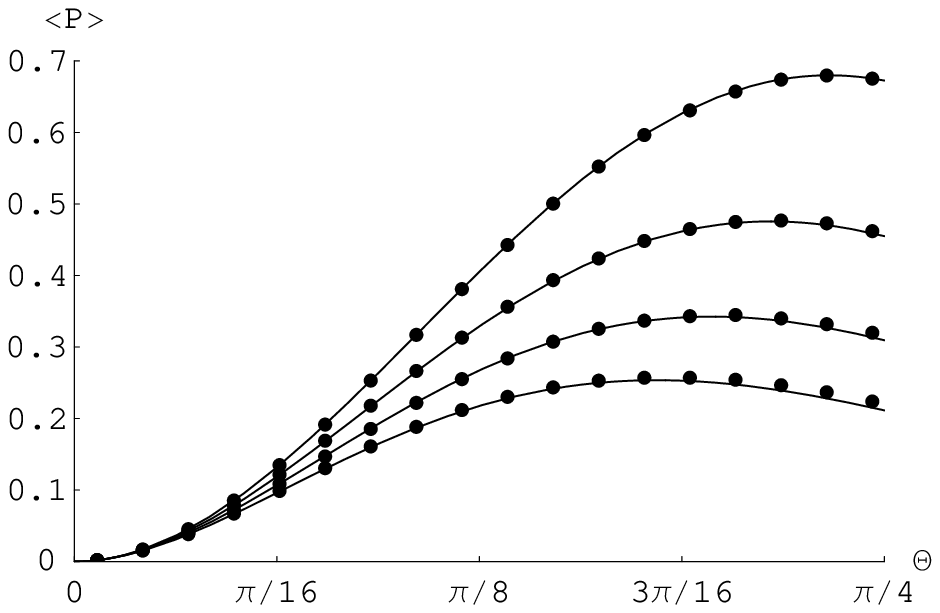}
\end{center}
\caption{$\bra P\ket(\Theta,x)$ as a function
of $\Theta$ (radian) with fixing $p$.
Both $\bra P \ket$ and $\Theta$ are dimensionless.
To estimate $\bra P\ket(\Theta,x)$, we put
$x=2\Theta(\arcsin\sqrt{1/2^{n}})^{-1}np$,
where $n=9$.
Four thin solid curves represent
$p=2.0\times 10^{-3}$, $4.0\times 10^{-3}$,
$6.0\times 10^{-3}$, and $8.0\times 10^{-3}$
in order from top to bottom.
Black circles represent results obtained by Monte Carlo simulations
of the $n=9$ cases (nine qubits).
Each circle is obtained for
$\Theta=(1/2)(2M+1)\theta$,
where $\theta=(\arcsin\sqrt{1/2^{n}})^{-1}$,
$n=9$, and $M\in\{0,1,...,M_{\mbox{\scriptsize max}}(=17)\}$.}
\lab{ptheta-fixx}
\end{figure}

Figure~\ref{px-fixtheta} shows
$\bra P\ket(\Theta,x)$ as a function
of $x$ with fixing $\Theta=\Theta_{\mbox{\scriptsize max}}$,
where
$\Theta_{\mbox{\scriptsize max}}=
(1/2)(2M_{\mbox{\scriptsize max}}+1)\theta$,
$M_{\mbox{\scriptsize max}}=17$,
$\theta=\arcsin\sqrt{1/2^{n}}$, and $n=9$.
(Hence, the independent parameter is only $p$ actually.)
At $x=0$, there is no error in the quantum process and
$\bra P \ket$ is nearly equal to unity.
As the error rate $x$ becomes larger,
$\bra P \ket$ decreases monotonically.

Figure~\ref{ptheta-fixx} shows
$\bra P\ket(\Theta,x)$ as a function
of $\Theta$ with fixing $p$.
Because we use the variable $x=2Mnp$ instead of $p$
in the perturbation theory,
we have to rewrite $x$ as
\beq
x=2Mnp=2\Theta(\arcsin\sqrt{1/2^{n}})^{-1}np,
\lab{parameter-x-with-fixing-p}
\eeq
which we obtain by substituting $\Theta=\lim_{n\rightarrow\infty}M\theta$
for $x=2Mnp$ without taking the limit $n\rightarrow\infty$,
and we give some finite $n$ to Eq.~(\ref{parameter-x-with-fixing-p}). 
In Fig.~\ref{ptheta-fixx}, we set $n=9$ and plot curves
with $p=2.0\times 10^{-3}$, $4.0\times 10^{-3}$,
$6.0\times 10^{-3}$, and $8.0\times 10^{-3}$
in order from top to bottom.
We also plot results of the simulations.
When we plot a result of the simulation for the $M$th step,
we put
\beq
\Theta=(1/2)(2M+1)\theta=(1/2)(2M+1)(\arcsin\sqrt{1/2^{n}})^{-1},
\lab{parameter-Theta-simulation-plot}
\eeq
where $n=9$ and $M\in\{0,1,...,M_{\mbox{\scriptsize max}}(=17)\}$.
We obtain Eq.~(\ref{parameter-Theta-simulation-plot})
from Eq.~(\ref{matrix-element-T0})
and $\Theta=\lim_{n\rightarrow\infty}M\theta$.

From Fig.~\ref{ptheta-fixx}, we notice that the maximum
value of $\bra P \ket$
is taken at $\Theta<\pi/4$ for each $p$
and the shift becomes larger as $p$ increases.
This fact means that
$\Theta_{\mbox{\scriptsize th}}
(p_{\mbox{\scriptsize c}},
P_{\mbox{\scriptsize th}})$
becomes smaller than $\pi/4$,
as $P_{\mbox{\scriptsize th}}$ decreases.
[We write
$\Theta_{\mbox{\scriptsize th}}(p,
P_{\mbox{\scriptsize th}})
=\lim_{n\rightarrow\infty}
M_{\mbox{\scriptsize th}}(p,
P_{\mbox{\scriptsize th}})\theta$
and
$M_{\mbox{\scriptsize th}}(p,
P_{\mbox{\scriptsize th}})$ represents the least number of the operations
iterated for amplifying the probability
of $|0\ket$ to
$P_{\mbox{\scriptsize th}}$
under the error rate $p$.]

Finally, we compute
$x_{\mbox{\scriptsize c}}$ as a function of $P_{\mbox{\scriptsize th}}$.
We show the result in Figs.~\ref{pthxc_linear} and \ref{pthxc_log}.
We obtain $x_{\mbox{\scriptsize c}}$
for $0\leq \forall P_{\mbox{\scriptsize th}}\leq 1$
as follows.
We calculate
$\tilde{\Theta}_{\mbox{\scriptsize th}}(x,P_{\mbox{\scriptsize th}})$
for given $P_{\mbox{\scriptsize th}}$
varying $x$ from zero,
where
$\tilde{\Theta}_{\mbox{\scriptsize th}}(x,P_{\mbox{\scriptsize th}})
=\lim_{n\rightarrow \infty}
\tilde{M}_{\mbox{\scriptsize th}}(x,P_{\mbox{\scriptsize th}})\theta$
and
$\tilde{M}_{\mbox{\scriptsize th}}(x,P_{\mbox{\scriptsize th}})$
represents the least number of the operations to amplify
the probability of $|0\ket$ to $P_{\mbox{\scriptsize th}}$
under given $x$.
(We use Newton's method for obtaining a root of $\Theta$
for the equation
$\bra P \ket(\Theta,x)
=P_{\mbox{\scriptsize th}}$
for given $x$.)
When $x$ becomes a certain value,
we cannot find a root of
$\tilde{\Theta}_{\mbox{\scriptsize th}}(x,P_{\mbox{\scriptsize th}})$
and we regard it as $x_{\mbox{\scriptsize c}}$.
By repeating this calculation
for many points of $P_{\mbox{\scriptsize th}}$,
we obtain the curve shown in Figs.~\ref{pthxc_linear} and \ref{pthxc_log}.

Using Eq.~(\ref{matrix-element-asymptotic-expansion}),
a tangent at $P_{\mbox{\scriptsize th}}=1$
is given by
\beq
x_{\mbox{\scriptsize c}}
=
c(1-P_{\mbox{\scriptsize th}}),
\quad\quad
c=-\frac{1}{C_{1}(\pi/4)}=\frac{8}{5},
\lab{tangent-line}
\eeq
because
$\tilde{\Theta}_{\mbox{\scriptsize th}}
(x_{\mbox{\scriptsize c}},P_{\mbox{\scriptsize th}})
=\pi/4$
and
$x_{\mbox{\scriptsize c}}=0$
for $P_{\mbox{\scriptsize th}}=1$.
This means that
the algorithm is effective
for
$2Mnp<(8/5)(1-P_{\mbox{\scriptsize th}})$
near $P_{\mbox{\scriptsize th}}=1$,
as shown in Fig.~\ref{pthxc_linear}.
This result is similar to a work
obtained by Bernstein and Vazirani \cite{Bernstein},
and Preskill \cite{Preskill},
as explained in Sec.~\ref{section-introduction}.
Moreover, we notice
$x_{\mbox{\scriptsize c}}>-(8/5)\log_{e}P_{\mbox{\scriptsize th}}$
near $P_{\mbox{\scriptsize th}}=0$ from Fig.~\ref{pthxc_log}.

\section{Discussions}
\lab{section-discussions}
From Fig.~\ref{pthxc_linear},
we find that the algorithm is effective for
$x=2Mnp<(8/5)(1-P_{\mbox{\scriptsize th}})$
near $P_{\mbox{\scriptsize th}}=1$,
and this relation is applied to a wide range of
$P_{\mbox{\scriptsize th}}$ approximately.
Thus, if we assume that $P_{\mbox{\scriptsize th}}$
is equal to a certain value
($1/2\leq P_{\mbox{\scriptsize th}}\leq 1$, for example),
we can expect that the algorithm works for
$x=2Mnp\leq O(1)$
($x$ is equal to or less than some constant.)
Hence, if the error rate $p$ is smaller
than an inverse of the number of quantum gates
$(2Mn)^{-1}$,
the algorithm is reliable.
If this observation holds good for other quantum algorithms,
it can serve as a strong foundation to realize quantum computation.

After we studied decoherence in Grover's algorithm
with a perturbation theory in Ref.~\cite{Azuma},
some other groups have tried similar analyses.
Shapira {\it et al}. investigate performance of Grover's algorithm
under unitary noise \cite{Shapira}.
They assume the noisy Hadamard gate
and estimate the success probability to detect a marked state
up to the first order perturbation.
Hasegawa and Yura consider decoherence
in the quantum counting algorithm,
which is a combination of Grover's algorithm and
the quantum Fourier transformation,
under the depolarizing channel \cite{Hasegawa}.

\bigskip
\noindent
{\bf \large Acknowledgment}
\smallskip

\noindent
The author thanks Osamu Hirota for encouragement.

\appendix
\section{Proof of Eq.~(\ref{F-h-Theta-expansion})}
\lab{appendix-proof-F-h-Theta}
In this section, we prove Eq.~(\ref{F-h-Theta-expansion}),
which we can rewrite in the form
\beq
f_{k}(\Theta)
=\mbox{Const.}\Theta^{k+2}+O(\Theta^{k+4})
\quad\quad
\mbox{for $k=0,1,...$}.
\lab{f-h-Theta-expansion}
\eeq
To put it more precisely,
we can obtain the following relations in which
Eq.~(\ref{f-h-Theta-expansion}) is included:
\beqa
f_{k}(\Theta)
&=&
a_{0}^{(k)}\Theta^{k+2}+a_{1}^{(k)}\Theta^{k+4}
+a_{2}^{(k)}\Theta^{k+6}+... \non \\
&=&
\sum_{j=0}^{\infty}a_{j}^{(k)}\Theta^{k+2(j+1)},
\lab{f-h-Theta-Taylor-series} \\
g_{k}(\Theta)
&=&
b_{0}^{(k)}\Theta^{k}+b_{1}^{(k)}\Theta^{k+2}
+b_{2}^{(k)}\Theta^{k+4}+... \non \\
&=&
\sum_{j=0}^{\infty}b_{j}^{(k)}\Theta^{k+2j}
\lab{g-h-Theta-Taylor-series} \\
&& \quad\quad
\mbox{for $k=0,1,...$}, \non
\eeqa
where $f_{k}(\Theta)$ and $g_{k}(\Theta)$
are defined in Eqs.~(\ref{f-0-Theta-definition}),
(\ref{g-0-Theta-definition}), (\ref{f-k-Theta-definition}),
and (\ref{g-k-Theta-definition}).

We prove Eqs.~(\ref{f-h-Theta-Taylor-series})
and (\ref{g-h-Theta-Taylor-series})
by mathematical induction.
First, when $k=0$,
we obtain the following results from Eqs.~(\ref{f-0-Theta-definition})
and (\ref{g-0-Theta-definition}):
\beqa
f_{0}(\Theta)
&=&
\sin^{2}2\Theta
=[\sum_{n=0}^{\infty}
\frac{(-1)^{n}2^{2n+1}}{(2n+1)!}\Theta^{2n+1}]^{2} \non \\
&=&
4\Theta^{2}-\frac{16}{3}\Theta^{4}
+\frac{128}{45}\Theta^{6}+..., \lab{f-0-Theta-Taylor-series} \\
g_{0}(\Theta)
&=&
\cos^{2}2\Theta
=[\sum_{n=0}^{\infty}
\frac{(-1)^{n}2^{2n}}{(2n)!}\Theta^{2n}]^{2} \non \\
&=&
1-4\Theta^{2}+\frac{16}{3}\Theta^{4}
+.... \lab{g-0-Theta-Taylor-series}
\eeqa
Thus, Eqs.~(\ref{f-h-Theta-Taylor-series})
and (\ref{g-h-Theta-Taylor-series}) are satisfied for $k=0$.

Next, assuming that Eqs.~(\ref{f-h-Theta-Taylor-series})
and (\ref{g-h-Theta-Taylor-series}) are satisfied for some $k$,
we investigate whether or not
Eqs.~(\ref{f-h-Theta-Taylor-series})
and (\ref{g-h-Theta-Taylor-series}) hold for $(k+1)$.
Let us consider Eq.~(\ref{f-h-Theta-Taylor-series})
for $(k+1)$.
From Eq.~(\ref{f-k-Theta-definition}), we obtain
\beq
f_{k+1}(\Theta)=
\int^{\Theta}_{0}d\phi
[f_{k}(\Theta-\phi)\cos^{2}2\phi
+g_{k}(\Theta-\phi)\sin^{2}2\phi]. \lab{f-(h+1)-Theta-definition}
\eeq
Here, we expand $\cos^{2}2\phi$ and $\sin^{2}2\phi$
as follows:
\beqa
\cos^{2}2\phi
&=&
\sum_{j=0}^{\infty}
c_{j}\phi^{2j}, \lab{cos2-2phi-Taylor-series} \\
\sin^{2}2\phi
&=&
\sum_{j=0}^{\infty}
d_{j}\phi^{2(j+1)}. \lab{sin2-2phi-Taylor-series}
\eeqa
From Eqs.~(\ref{f-h-Theta-Taylor-series}),
(\ref{g-h-Theta-Taylor-series}),
(\ref{cos2-2phi-Taylor-series}), and
(\ref{sin2-2phi-Taylor-series}),
we can rewrite Eq.~(\ref{f-(h+1)-Theta-definition})
in the form
\beqa
f_{k+1}(\Theta)
&=&
\int^{\Theta}_{0}d\phi
\sum_{i=0}^{\infty}
\sum_{j=0}^{\infty}
[a_{i}^{(k)}c_{j}(\Theta-\phi)^{k+2(i+1)}\phi^{2j} \non \\
&&\quad\quad\quad\quad
+b_{i}^{(k)}d_{j}(\Theta-\phi)^{k+2i}\phi^{2(j+1)}].
\lab{f-(h+1)-Theta-integral-summation}
\eeqa
Applying the following formula
\beq
\int^{\Theta}_{0}d\phi
(\Theta-\phi)^{i}\phi^{2j} 
=\frac{i!(2j)!}{(i+2j+1)!}\Theta^{i+2j+1}
\lab{formula-integral}
\eeq
to Eq.~(\ref{f-(h+1)-Theta-integral-summation}),
we find that $f_{k+1}(\Theta)$
includes only terms of $\Theta^{k+3}$,
$\Theta^{k+5}$, $\Theta^{k+7}$, ....
Therefore, Eq.~(\ref{f-h-Theta-Taylor-series})
holds for $(k+1)$.
Next, let us consider Eq.~(\ref{g-h-Theta-Taylor-series})
for $(k+1)$.
From Eq.~(\ref{g-k-Theta-definition}), we obtain
\beq
g_{k+1}(\Theta)=
\int^{\Theta}_{0}d\phi
[g_{k}(\Theta-\phi)\cos^{2}2\phi
+f_{k}(\Theta-\phi)\sin^{2}2\phi].
\lab{g-(h+1)-Theta-definition}
\eeq
Using Eqs.~(\ref{f-h-Theta-Taylor-series}),
(\ref{g-h-Theta-Taylor-series}),
(\ref{cos2-2phi-Taylor-series}), and
(\ref{sin2-2phi-Taylor-series}),
we can rewrite Eq.~(\ref{g-(h+1)-Theta-definition})
in the form
\beqa
g_{k+1}(\Theta)
&=&
\int^{\Theta}_{0}d\phi
\sum_{i=0}^{\infty}
\sum_{j=0}^{\infty}
[b_{i}^{(k)}c_{j}(\Theta-\phi)^{k+2i}\phi^{2j} \non \\
&&\quad\quad\quad\quad
+a_{i}^{(k)}d_{j}(\Theta-\phi)^{k+2(i+1)}\phi^{2(j+1)}].
\lab{g-(h+1)-Theta-integral-summation}
\eeqa
Applying Eq.~(\ref{formula-integral})
to Eq.~(\ref{g-(h+1)-Theta-integral-summation}),
we find that $g_{k+1}(\Theta)$
includes only terms of $\Theta^{k+1}$,
$\Theta^{k+3}$, $\Theta^{k+5}$, ....
Therefore, Eq.~(\ref{g-h-Theta-Taylor-series})
holds for $(k+1)$.
Hence, from mathematical induction,
we conclude that
Eqs.~(\ref{f-h-Theta-Taylor-series}) and
(\ref{g-h-Theta-Taylor-series})
are satisfied for $k=0,1,...$.
This implies that Eq.~(\ref{f-h-Theta-expansion}) holds.

\end{document}